\pdfoutput=1
\documentclass[twocolumn]{emulateapj}

\usepackage{latexsym}
\usepackage{gensymb}
\usepackage{amsmath}
\usepackage{graphicx}
\usepackage{color}
\usepackage[colorlinks]{hyperref}
\hypersetup{
    colorlinks,	
    citecolor=blue,
}

\makeatletter

\newcommand{\Rmnum}[1]{\expandafter\@slowromancap\romannumeral #1@}
\makeatother

\shorttitle{Testing the Kerr metric with Fairall~9}
\shortauthors{Liu et al.}

\begin{document}

\title{Reflection features in the X-ray spectrum of Fairall~9 and implications for tests of general relativity}

\author{Honghui~Liu\altaffilmark{1}, Haiyang~Wang\altaffilmark{1}, Askar~B.~Abdikamalov\altaffilmark{1}, Dimitry~Ayzenberg\altaffilmark{1}, and Cosimo~Bambi\altaffilmark{1,\dag}}

\altaffiltext{1}{Center for Field Theory and Particle Physics and Department of Physics, 
Fudan University, 200438 Shanghai, China. \email[\dag E-mail: ]{bambi@fudan.edu.cn}}

\begin{abstract}
X-ray reflection spectroscopy is potentially a powerful tool to probe the spacetime geometry around astrophysical black holes and test general relativity in the strong field regime. However, precision tests of general relativity are only possible if we employ the correct astrophysical model and we can limit the systematic uncertainties. It is thus crucial to select the sources and the observations most suitable for these tests. In this work, we analyze simultaneous observations of \textit{XMM-Newton} and \textit{NuSTAR} of the supermassive black hole in Fairall~9. This source has a number of properties that make it a promising candidate for tests of general relativity using X-ray reflection spectroscopy. Nevertheless, we find that with the available data there is not a unique interpretation of the spectrum of Fairall~9, which prevents, for the moment, to use this source for robust tests of general relativity. This issue may be solved by future X-ray missions with a higher energy resolution near the iron line.  
\end{abstract}



\section{introduction}

Einstein's theory of general relativity is a pillar in modern physics. However, its predictions have been mainly tested in weak gravitational fields, while the strong field regime is still largely unexplored~\citep{2014LRR....17....4W}. Astrophysical black holes are ideal laboratories for testing general relativity in the strong field regime and present and near future observational facilities are promising to provide unprecedented high quality data to use these objects for testing fundamental physics~\citep{2019arXiv190603871B}.

In 4-dimensional general relativity, uncharged black holes are described by the Kerr solution and are completely characterized by their mass and spin angular momentum~\citep{1963PhRvL..11..237K}. This is the celebrated result of the no-hair theorem, which holds under specific assumptions~\citep{1971PhRvL..26..331C,1975PhRvL..34..905,2012LRR....15....7C}. The spacetime metric around astrophysical black holes is thought to be well approximated by the Kerr solution~\citep{2014PhRvD..89l7302B,2018AnP...53000430B}. However, macroscopic deviations from the Kerr metric are possible in the presence of exotic matter~\citep{2014PhRvL.112v1101H}, macroscopic quantum gravity effects~\citep{2011arXiv1112.3359D,2017NatAs...1E..67G,2019arXiv191111200C}, and if general relativity is not the correct theory of gravity~\citep{2009PhRvD..79h4043Y,2014PhRvD..90d4066A}. Testing the Kerr metric around astrophysical black holes is a test of general relativity in the strong field regime and can be seen as the counterpart of the tests of the Schwarzschild solution in the weak field limit with Solar System experiments.

The study of the properties of the electromagnetic radiation emitted by stars or gas orbiting an astrophysical black hole can test the Kerr black hole hypothesis~\citep{2011MPLA...26.2453B,2016CQGra..33l4001J,2017RvMP...89b5001B,2018GReGr..50..100K}. Among all the electromagnetic techniques for testing the Kerr nature of an astrophysical black hole, X-ray reflection spectroscopy~\citep{2000PASP..112.1145F,2006ApJ...652.1028B,2013mams.book.....B,2014SSRv..183..277R,Walton2013} is the most mature method and currently the only one capable of providing quantitative constraints on possible deviations from general relativity in the strong gravity region of black holes~\citep{2018PhRvL.120e1101C,2018PhRvD..98b3018T,2019ApJ...874..135T,2019ApJ...875...56T,2018ApJ...865..134X,2019ApJ...884..147Z}.

In the past few years, our group has developed the relativistic reflection model \texttt{relxill\_nk}~\citep{Bambi2017,2019ApJ...878...91A}, which is an extension of the \texttt{relxill} package~\citep{2013MNRAS.430.1694D,Garcia2013,Garcia2014} to non-Kerr spacetimes. The model employs a background metric more general than the Kerr solution and that includes the Kerr solution as a special case. While the model has been used with different black hole spacetimes, the key-idea is the presence of a number of ``deformation parameters'' that are introduced to quantify possible deviations from the Kerr geometry. These deformation parameters are treated as free parameters during the fits and then we can check {\it a posteriori} if observational data require vanishing deformation parameters, namely observations can be explained better if the spacetime metric around black holes is described by the Kerr metric and we can constrain deviations from the Kerr geometry.

Precision tests of general relativity using X-ray reflection spectroscopy -- as well as any other technique -- are only possible if we can limit the systematic uncertainties, so that the measurements are not only precise but even accurate. In general, relativistic reflection models have a number of simplifications that introduce systematic uncertainties in the final measurements and this may indeed prevent robust tests of general relativity~\citep{2019PhRvD..99l3007L}. While we can surely work to improve our theoretical models, it is also crucial to select the most suitable sources and observations.

In the present study, we analyze simultaneous observations of \textit{XMM-Newton} and \textit{NuSTAR} of the supermassive black hole in Fairall~9. Such a source has a number of properties suggesting that it is potentially quite a promising target for our tests of the Kerr black hole hypothesis using X-ray reflection spectroscopy. More specifically, these properties are:
\begin{enumerate}
\item The Eddington-scaled bolometric luminosity of the source is around 5\%~\citep{2009MNRAS.392.1124V,2013mams.book.....B}, so the accretion disk should be described well by the Novikov-Thorne model with the inner edge at the innermost stable circular orbit (ISCO)~\citep{2010ApJ...718L.117S,2010MNRAS.408..752P,2011MNRAS.414.1183K}. Note that standard relativistic reflection models employing Novikov-Thorne disks are commonly used even to fit data of sources with an Eddington-scaled bolometric luminosity exceeding 30\%, see e.g. Tab.~1 in \citet{2013mams.book.....B}, but such measurements may not be accurate~\citep{2020MNRAS.491..417R,2019arXiv191106605R}.
\item The inner edge of the accretion disk was found very close to the black hole in~\citet{Lohfink2016}. This maximizes the relativistic effects in the spectrum and is particularly useful in a test of general relativity.
\item The source is an AGN, so its accretion disk is cold, which is the approximation used in the calculations of the reflection spectrum in our model; see~\citet{Garcia2013,Garcia2014}. In the case of black hole binaries, the temperature of the inner part of the accretion disk can be up to about 1~keV, where the validity of our model is questionable.
\item There are simultaneous observations of \textit{XMM-Newton} and \textit{NuSTAR}; the former guarantees a good energy resolution near the iron line and the latter provides data up to 80~keV.
\item The source is one of the so-called ``bare'' AGNs, namely there is no warm absorber along the line of sight~\citep{Emmanoulopoulos2011}. This means we do not have to worry about the astrophysical uncertainties related to a warm absorber, which is useful in a test of general relativity.
\item The source is not very variable, suggesting no changes in the geometry of the accretion flow around the black hole~\citep{Lohfink2016}. This is useful because we can assume that the coronal geometry, and therefore the disk intensity profile, do not change over the observation and we can fit the data with a single emissivity profile.
\item Fairall~9 is a bright source.
\end{enumerate}

All these properties make Fairall~9 quite a promising source for robust tests of general relativity using X-ray reflection spectroscopy. 

The rest of the paper is organized as follows. In Section~\ref{s-obs}, we present the observational data. In Section~\ref{s-ana}, we report our spectral analysis, first assuming the Kerr metric and employing \texttt{relxill}, then testing the Kerr metric with \texttt{relxill\_nk}. Discussion and conclusions are in Section~\ref{s-dis}.

\begin{table}[!hb]
\renewcommand\arraystretch{1.8} 
\centering
\caption{\label{studies}}
\begin{tabular}{ccp{2.8cm}p{1.5cm}} 
\hline\hline
Spin ($a_*$) & Incl ($i$) [deg] & Reference & Satellite \\
\hline
$0.60_{-0.07}^{+0.07}$ & $44_{-1}^{+1}$ & \cite{Schmoll2009} & \textit{Suzaku}\\
$0.39_{-0.30}^{+0.48}$ & $64_{-9}^{+7}$ & \cite{Emmanoulopoulos2011} & \textit{XMM} \\
$0.67_{-0.11}^{+0.10}$ & $33_{-3}^{+3}$ & \cite{Patrick2011} & \textit{Suzaku}\\
$>0.64$ & $45_{-9}^{+13}$ & \cite{Walton2013} & \textit{Suzaku}\\
$>0.997$ & $<11$ & \cite{Lohfink2016} & \textit{XMM} \& \textit{NuSTAR}\\
\hline
\vspace{-0.2cm}
\tablenotetext{0}{Previous measurements of the spin parameter and inclination of the accretion disk of Fairall 9.}
\end{tabular}
\end{table}

\section{observation data}\label{s-obs}

Fairall 9 is a luminous Seyfert 1 galaxy ($z=0.047$) in the center of which lies a supermassive black hole: $(2.55 \pm 0.56)\times 10^8 M_{\odot}$~\citep{Peterson2004}. The X-ray spectrum of Fairall 9 is know for lack of warm absorber~\citep{Emmanoulopoulos2011}, which enable us to directly probe the physical properties of its central black hole. With little contamination from the line-of-sight absorption material, the X-ray spectra of Fairall 9 have been well studied in the last decades. Measurements of the black hole spin are summarized in Tab.~\ref{studies}. \citet{Lohfink2012} also studied several \textit{Suzaku} and \textit{XMM-Newton} data sets of Fairall 9, but found that the measured disk parameters are model dependent. It's only in \citet{Lohfink2016} that data from both \textit{XMM-Newton} and \textit{NuSTAR} were available and the authors found an extreme high spin of the black hole. However, in the same year, \citet{Yaqoob2016} reported that there is no spin signature in the X-ray spectrum of Fairall 9 with data set SUZA in Tab.~\ref{summary}.

\subsection{Available spectral data sets}

There are several observations of Fairall 9 by \textit{Suzaku}, \textit{XMM-Newton} and \textit{NuSTAR}. Tab.~\ref{summary} gives a summary of these observations. Throughout this work, the spectral modeling is focused on the last \textit{XMM} observation (XMME) and the simultaneous \textit{NuSTAR} observation. 

\begin{table*}
    \renewcommand\arraystretch{1.5}
    \centering
    \caption{\label{summary}}
    \begin{tabular}{ccccc}
        \hline\hline
        Satellite & Obs Date & ObsID & Exposure (ks) & Reference name \\
        \textit{XMM} & 2000-07-05 & 0101040201 & 26 & XMMA \\
        \textit{Suzaku} & 2007-06-07 & 702043010 & 140 & SUZA \\
        \textit{XMM} & 2009-12-09 & 0605800401 & 91 & XMMB \\
        \textit{Suzaku} & 2010-05-09 & 705063010 & 143 & SUZB \\
        \textit{XMM} & 2013-12-19 & 0720000101 & 42 & XMMC \\
        \textit{XMM} & 2014-01-02 & 0721110201 & 32 & XMMD \\
        \textit{XMM} & 2014-05-09 & 0741330101 & 82 & XMME \\
        \textit{NuSTAR} & 2014-05-09 & 60001130002 \& 60001130003 & 143 & NU \\
        \hline
        \vspace{-0.2cm}
        \tablenotetext{0}{Summary of available data of Fairall 9 from \textit{Suzaku}, \textit{XMM-Newton} and \textit{NuSTAR}. The reference names follow the convention in~\citet{Lohfink2016}.} 
    \end{tabular}
\end{table*}

\subsection{\textit{XMM-Newton}}
For \textit{XMM-Newton}~\citep{2001A&A...365L...1J}, we only use EPIC-pn data to do spectral modeling. The data is reduced with the \textit{XMM-Newton} Science Analysis System (SAS) version 18.0.0 and the latest calibration files until 2019 Oct. The spectra are extracted using tool \texttt{evselect} with default pattern (single and double events) selected. We extract the source spectra from a circular region with radius of 32.5 arcsec centered on the source. The background spectra are taken from a circular region with the same size near the source. We then use tasks \texttt{rmfgen} and \texttt{arfgen} to produce response files. The pn spectra are rebinned to ensure a minimal signal to noise ratio of 10. We also include a \texttt{gainshift} to account for the gain problem of the data as noted in~\citet{Marinucci2014}.

\subsection{NuSTAR}
The \textit{NuSTAR} satellite~\citep{Harrison} observed Fairall 9 on 2014 May 09. We use \texttt{nupipeline} v0.4.6 and the 20190812 version of NuSTAR CALDB to produce clean event files. The circular source region with a radius of 60 arcsec and the background region with the same size are created with \texttt{ds9}. We then use the \texttt{nuproducts} task to extract source and background spectra. Since there are two consecutive observations, we use the task \texttt{addascaspec} to combine the two spectra of each detector. The spectra are rebinned to a minimal counts of 50 for the fidelity of Chi-squared statistics.

\section{spectral analysis}\label{s-ana}

\begin{figure}
    \centering
    \includegraphics[width=8cm, trim={0.0cm 0.0cm 0.0cm 0.0cm}]{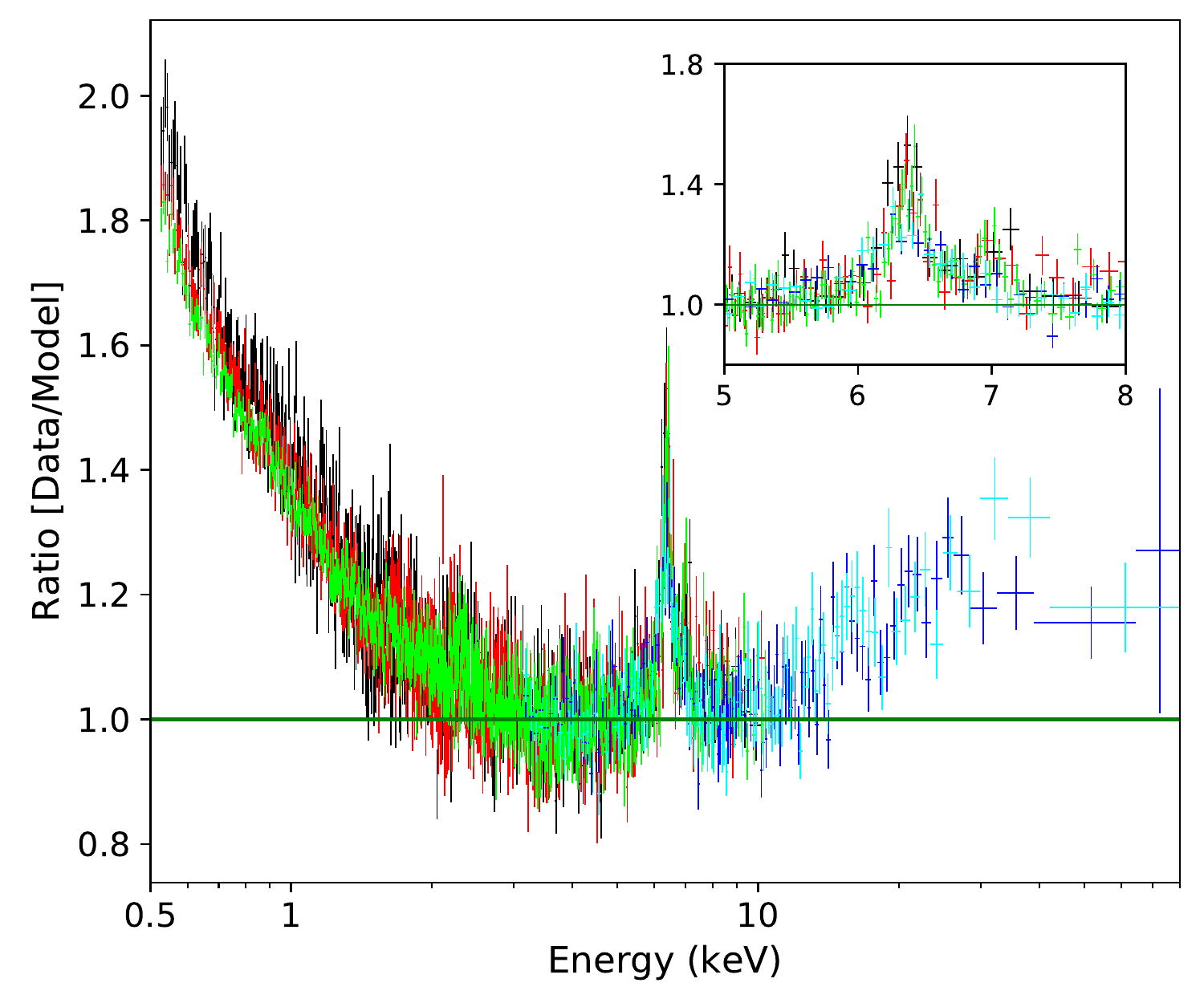}
    \caption{Data/Model residuals to a simple absorbed cutoff power-law model for observations XMMC (black), XMMD (red), XMME (green), FPMA (blue) and FPMB (cyan). The spectra are first fitted by ignoring data below 2.5 keV, 6--8 keV and above 10 keV. Then the ignored data are included to make this plot. The energy is given in the AGN frame. \label{ratios}}
\end{figure}

We use XSPEC v12.10.1f~\citep{xspec} and $\chi^2$ statistics to model the simultaneous \textit{XMM} and \textit{NuSTAR} spectra. We set the solar element abundances to~\citet{Wilms2000} and cross sections to~\citet{Verner1996}. To describe Galactic neutral absorption, model \texttt{Tbabs} with the column density ($n_{\rm H}$) fixed at $3.15 \times 10^{20} {\rm cm}^{-2}$ is always included in spectral modeling. To reveal reflection features in the X-ray spectra, we first fit the data with a simple absorbed power-law model which can be written in XSPEC notation as \texttt{Tbabs*zCutoffpl}. The best-fit data to model ratios are shown in Fig.~\ref{ratios}. It is clear that there is a strong soft excess below 2 keV and a Compton hump peaking at 30 keV. Fig.~\ref{ratios} also shows a prominent narrow iron ${\rm K}{\alpha}$ emission line at 6.4 keV in the rest frame and the associated ${\rm K}{\beta}$ line around 7.0 keV. These narrow line features together with the Compton hump indicate the existence of non-relativistic cold reflection in the source. It is also remarkable that the residuals are similar for three \textit{XMM} observations even though their overall spectra are rather different (see Figure. 2 in \citet{Lohfink2016}). This may indicate a stable corona disk geometry.

\begin{figure}
    \centering
    \includegraphics[width=8cm, trim={0.0 0.0 0.0 0.0}]{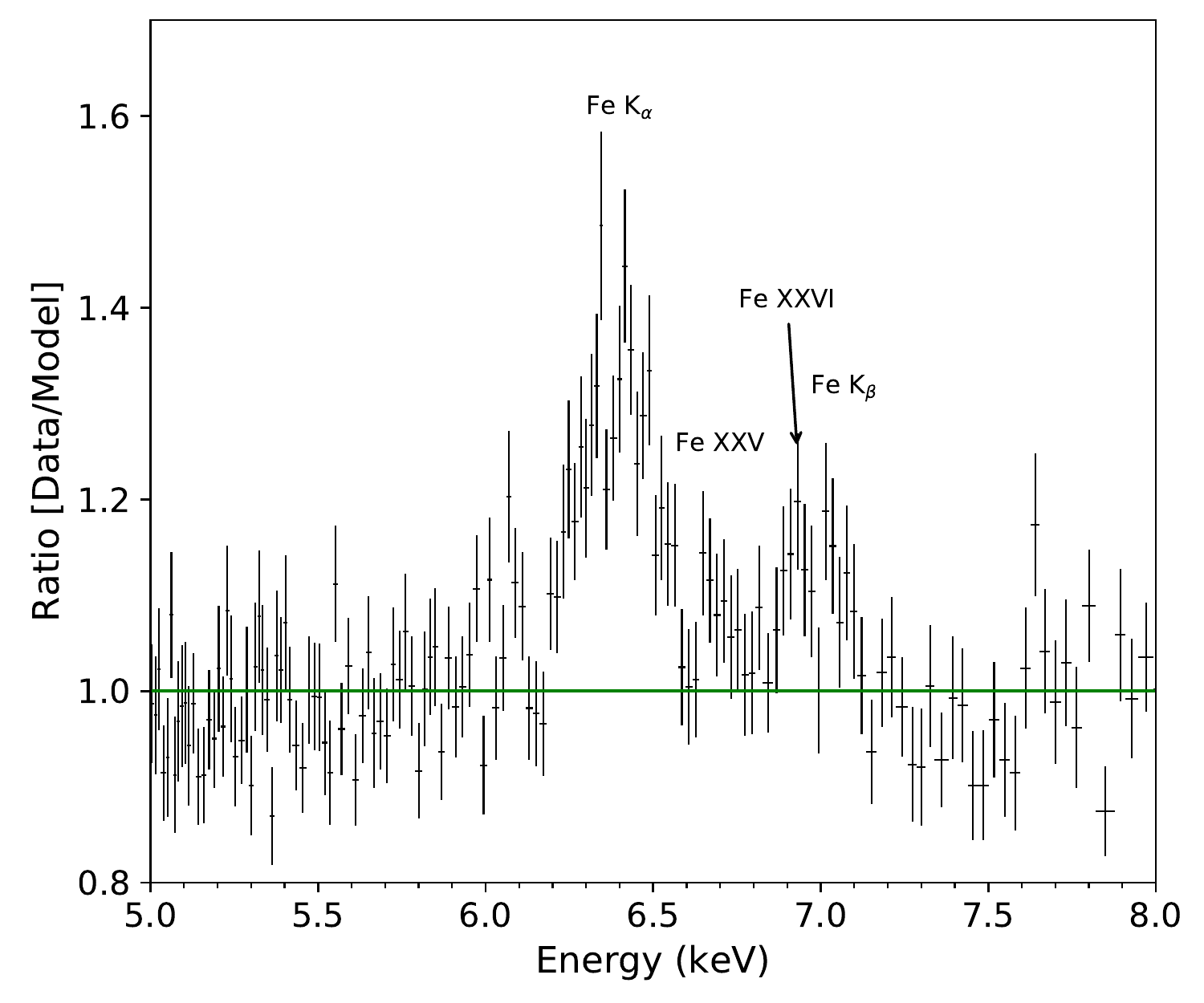}
    \caption{Residuals (in the rest frame) to a simple power-law model for data of XMME. The spectrum is fitted in the energy range 2.5-10 keV. Clear detection of line features is present. The residuals are grouped only for plotting purpose. \label{xmme-ratio}}
\end{figure}

A detailed examination of the residuals in Fig.~\ref{xmme-ratio} for XMME data shows a line feature between the ${\rm Fe} ~{\rm K}{\alpha}$ and the ${\rm Fe} ~ {\rm K}{\beta}$ peaks. This line feature could be emission from Fe \textsc{xxv} or the blue wing of broad iron line. The peak at 7 keV in Fig.~\ref{xmme-ratio} is actually a composition of  Fe \textsc{xxvi} and Fe ${\rm K}{\beta}$ lines. We include two narrow ($\sigma = 10$ eV) gaussian profiles to model these narrow emission lines from highly ionized iron (6.7 keV and 6.96 keV for Fe \textsc{xxv} and \textsc{xxvi} respectively) as done in \citet{Walton2013} and \citet{Patrick2011}.

The soft excess feature below 2 keV (see Fig.~\ref{ratios}) is commonly seen in X-ray spectra of Seyfert AGNs~\citep{Bianchi2009, Scott2012}. The origin of this soft excess component is still under debate. One of the two major explanations is the Comptonization scenario~\citep{Jin2009, Matt2014}. Photons from the disk are Comptonized by a warm corona which is cooler and optically thicker than the hot corona that produces the primary X-ray emission. The other explanation treats the soft excess as a signature of blurred ionized reflection~\citep{Crummy2006, Walton2013}. Sometimes both scenarios can explain the same data set equally well~\citep{Garcia2019}. We also try both scenarios to model the soft excess of Fairall 9.

\begin{figure}
    \centering
    \includegraphics[width=0.47\textwidth, trim={0.0cm 0.0cm 0.0cm 0.0cm},clip]{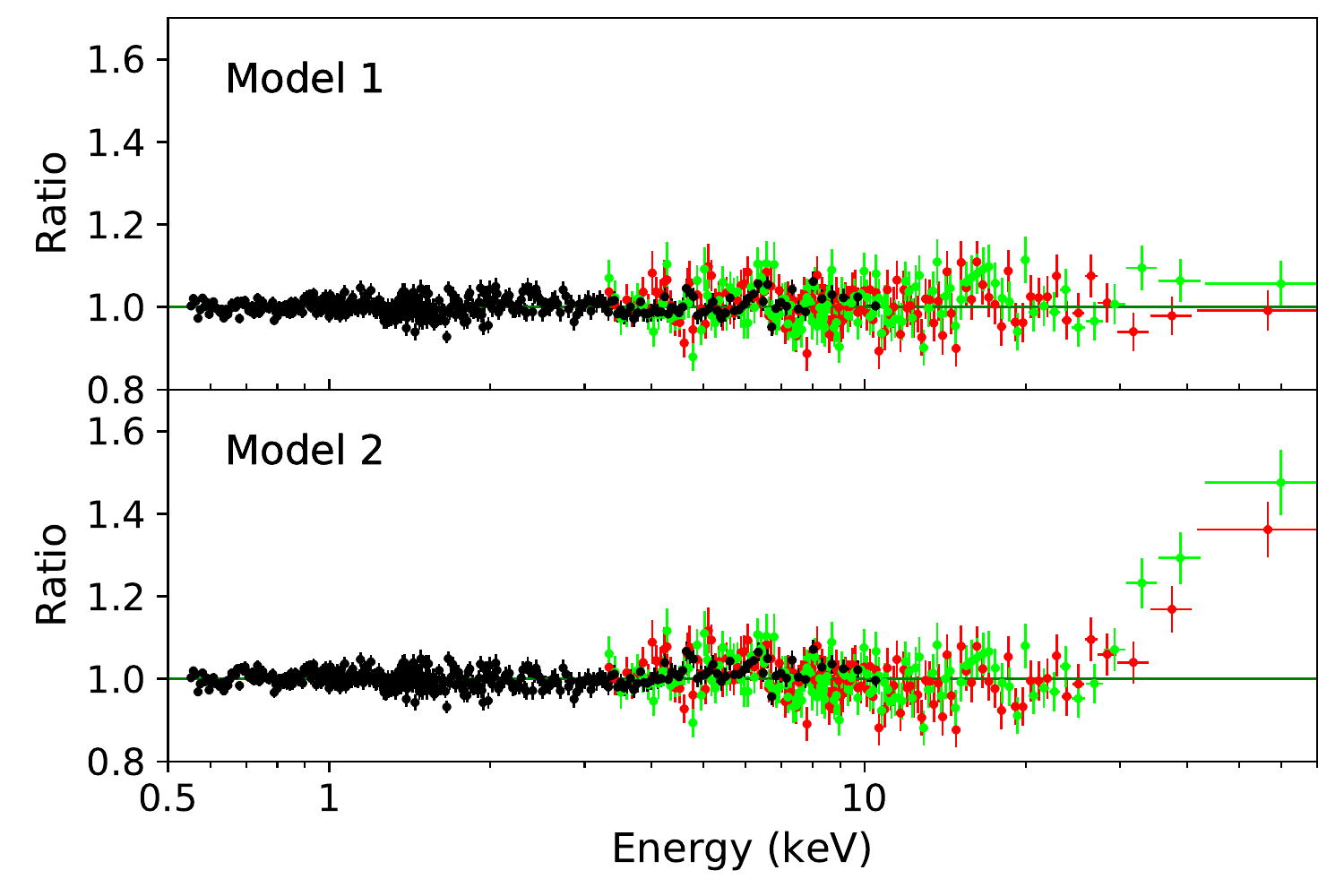}
    \caption{Residuals (in the rest frame) for the best fits with Model 1 and Model 2. The data shown are XMME (black), FPMA (Red) and FPMB (green). \label{m12-ratio}}
\end{figure}

\textit{Model 1.} We include model \texttt{nthComp}~\citep{Zdziarski1996, Zycki1999} to model the soft excess and a  cutoff power-law component for the primary continuum. For the cold reflection component, we model it with \texttt{xillver}~\citep{Garcia2013}. The ionization parameter ($\log\xi$) is set to 0 as expected for cold reflection. The inclination is fixed at $3^\circ$ for being not sensitive to the fit. We set the reflection fraction of \texttt{xillver} to $-1$ so it returns only the reflected component. The power-law index $\Gamma$ and cutoff energy $E_{\rm cut}$ are tied between \texttt{xillver} and \texttt{zCutoffpl}. The best-fit parameters are shown in Tab.~\ref{fits1} and data to model residuals in Fig.~\ref{m12-ratio}. Without including any blurred reflection component, we find this simple model already gives a good fit ($\chi^2_{\nu} = 1.04$) to the data. There is also no significant unresolved features left in the residuals. We have also tried this non-relativistic model to the data set XMMB, XMMC, XMMD and SUZA in Tab.~\ref{summary} separately and we always get good fits. Data from XMMA and SUZB are ignored for their problem noted in~\cite{Lohfink2012}.

\textit{Model 2.} Ionized blurred reflection from the inner part of the accretion disk is also a possible explanation of soft excess. We implement the model \texttt{relxill} v1.2.0~\citep{Garcia2014} to fit both the soft excess and possible relativistic reflection component. During the fit, the radial intensity profile of the incident radiation is assumed to be a power-law. We also tried broken power-law configuration but it does not make significant difference to the statistics or best-fit values of other parameters. The incident radiation to the accretion disk is presumed to be the primary emission. The disk inclination $i$ and iron abundance $A_{\rm Fe}$ are linked between \texttt{xillver} and \texttt{relxill}. We assume that the inner radius of the accretion disk is at the ISCO. From the best-fit values in Tab.~\ref{fits1}, we find that with two more free parameters, this model gives a $\chi^2$ larger (more than 150) than \textit{Model 1}. The residuals in Fig.~\ref{m12-ratio} also shows a clear hard excess in the \textit{NuSTAR} spectra above 20 keV as found by~\citet{Lohfink2016}. The acceptable fit in low energy band is expected since the fit is dominated by data below 10 keV. We also tried to fit the data with model \texttt{relxillD} (and the associated \texttt{xillverD}~\citep{Garcia2016}) in which the disk electron density is variable. However, this model does not improve the fit and the disk density is tightly constrained near the value assumed by \texttt{relxill} ($n_{\rm e} = 10^{15}~{\rm cm}^3$). We thus conclude that only primary emission, cold reflection and blurred reflection are not sufficient to model the spectra of Fairall 9. An additional Comptonization component is necessary. This result also addresses the importance of including broad band data when doing spectral analysis. If we do not have data above 20 keV, then both \textit{Model 1} and \textit{Model 2} can very well fit the spectra. To demonstrate this point, \textit{Model 2} is also used to fit the data from XMMB, XMMC, XMMD and SUZA. The best fit statistics are comparable to what we get with \textit{Model 1} with no significant features left.

\begin{table*}
    \centering
    \caption{Best-fit values for the two scenarios to model the soft excess. \label{fits1}}
    \renewcommand\arraystretch{1.8}
    \begin{tabular}{lcccc}
        \hline\hline
        Description & Component & Parameter & Model 1 & Model 2 \\
        \hline
        Galactic absorption & \texttt{Tbabs} & $N_{\rm H} (10^{20} {\rm cm}^2)$ & $3.15^*$ & $3.15^*$ \\
        \hline
        Soft excess & \texttt{nthComp} & $\Gamma$ & $2.82_{-0.04}^{+0.04}$ & - \\
        & \texttt{nthComp} & $z$ & $0.047^*$ & -  \\
        & \texttt{nthComp} & $kT_{\rm e}$ (keV) & $0.49_{-0.07}^{+0.08}$& - \\
        & \texttt{nthComp} & Norm ($10^{-3}$) & $2.6_{-0.3}^{+0.3}$ & -\\
        \hline
        Cold reflection & \texttt{xillver} & $A_{\rm Fe}$ & $3.1_{-0.7}^{+0.9}$ & $0.70_{-0.08}^{+0.06}$ \\
        & \texttt{xillver} & Norm ($10^{-5}$) & $4.8_{-0.8}^{+0.8}$ & $16.0_{-1.3}^{+0.7}$\\
        \hline
        Ionized reflection & \texttt{relxill} & $q_{\rm in} = q_{\rm out}$ & - & $8.6_{-1.3}^{+0.9}$ \\
        & \texttt{relxill} & $a_*$ & - & $0.955_{-0.007}^{+0.011}$ \\
        & \texttt{relxill} & $i$ (deg) & - & $<14$ \\
        & \texttt{relxill} & $\log\xi$ & - & $2.70_{-0.10}^{+0.02}$ \\
        & \texttt{relxill} & Norm ($10^{-5}$) & - & $6.6_{-0.4}^{+0.5}$  \\
        \hline
        Continuum & \texttt{zCutoffpl} & $\Gamma$ & $1.86_{-0.03}^{+0.03}$ & $2.189_{-0.013}^{+0.005}$ \\
        & \texttt{zCutoffpl} & $E_{\rm cut}$ (keV) & $282_{-86}^{+196}$ & $>930$ \\
        & \texttt{zCutoffpl} & Norm ($10^{-3}$) & $7.3_{-0.3}^{+0.3}$ & $8.67_{-0.18}^{+0.08}$ \\
        \hline
        &  $\chi^2/\nu$ & & 2222.96/2140 & 2387.94/2138 \\
        & $\chi^{2}_{\nu}$ & & 1.04 &  1.12\\
        \hline
        \tablenotetext{0}{\textbf{Note.} Model 1: \texttt{Tbabs*(nthComp + xillver + zCutoffpl)}. Model 2: \texttt{Tbabs*(relxill + xillver + zCutoffpl)}. The parameters with $^*$ are fixed at given values in the fit. Uncertainties are given at 90\% confidence level.}
    \end{tabular}
\end{table*}

\begin{figure}
    \centering
    \includegraphics[width=0.47\textwidth, trim={0.0cm 0.0cm 0.0cm 0.0cm},clip]{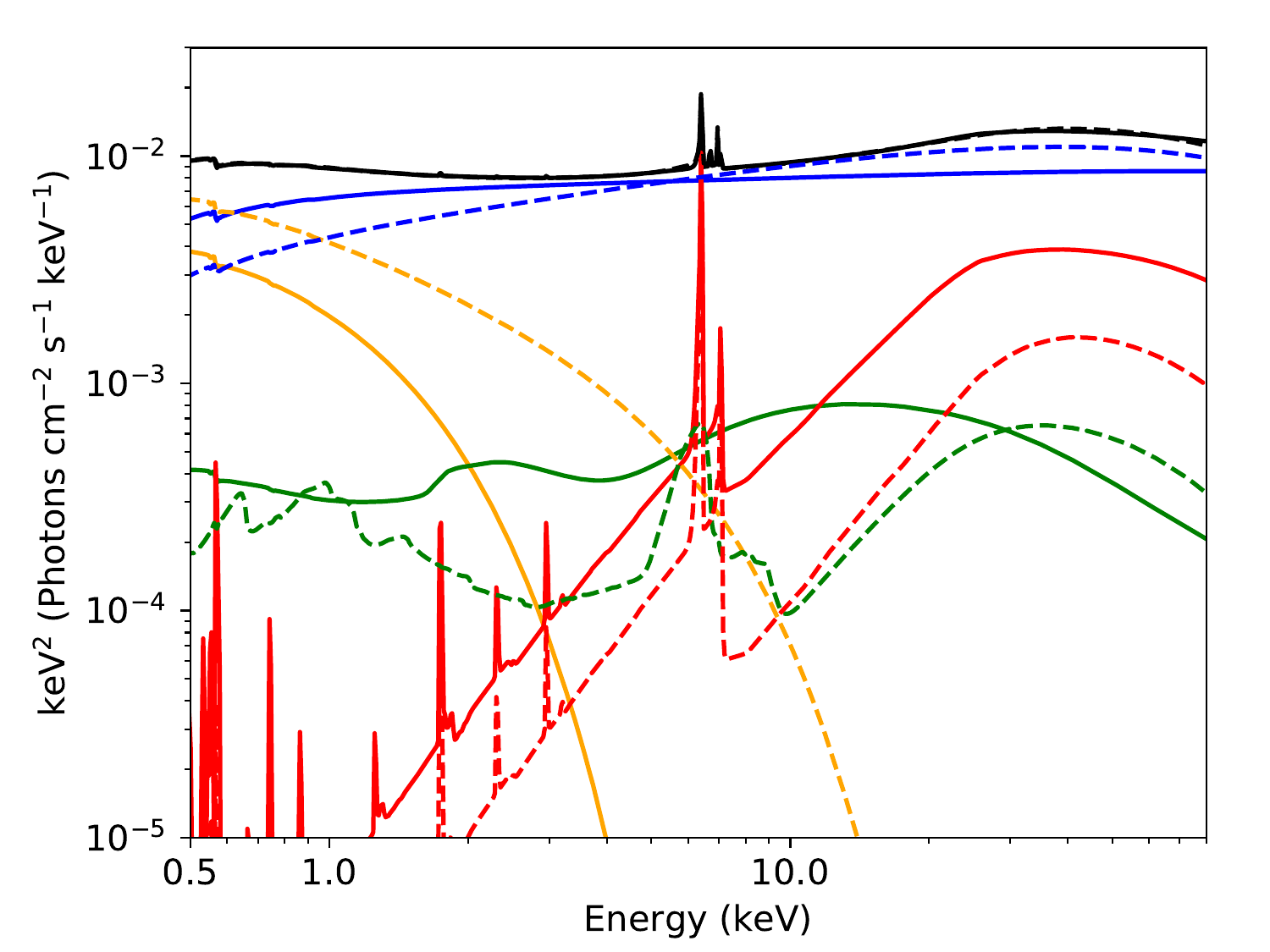}
    \caption{Components for the best fit of Model 3.1 (solid lines) and Model 3.2 (dashed lines). The colors correspond to: the total model (black), primary emission (blue), warm corona emission (orange), blurred reflection (green) and cold reflection (red). The emission lines from highly ionized iron are omitted for clarity. \label{models}}
\end{figure}

\begin{figure}
    \centering
    \includegraphics[width=0.47\textwidth, trim={0.0cm 0.0cm 0.0cm 0.0cm},clip]{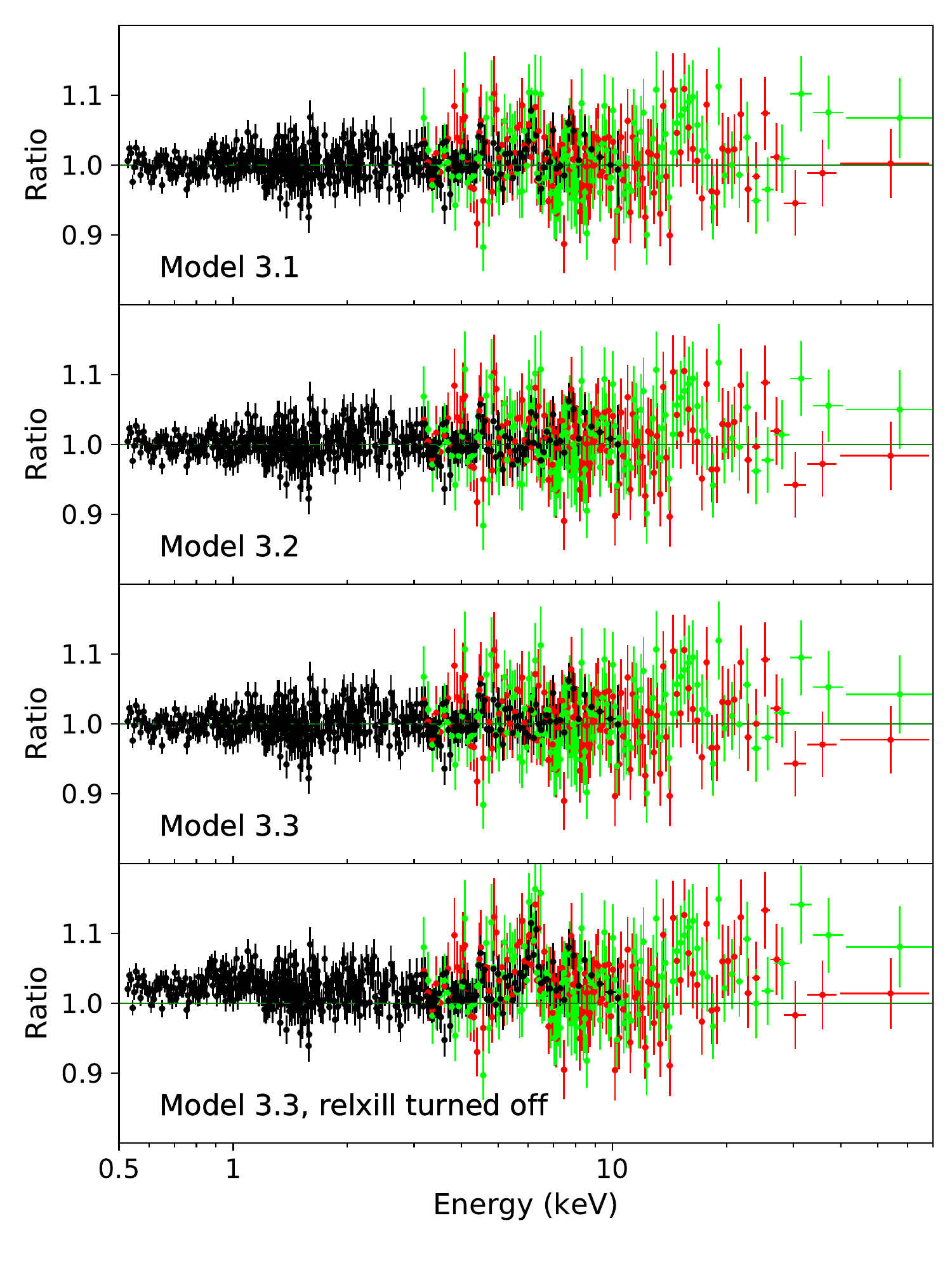}
    \caption{Data to best-fit model ratios for the 3 configurations of Model 3 and for Model 3.3 with relativistic reflection component turned off. The energy is given in the AGN rest frame. \label{ratios-Kerr}}
\end{figure}

\textit{Model 3.} Although we see from the top panel of Fig~\ref{m12-ratio} that \textit{Model 1} already fits the Fairall 9 spectra in a good manner, it is still interesting to know how adding a blurred reflection component can improve the fit. So we add a \texttt{relxill} component to \textit{Model 1}. The total model is similar to what is used in \citet{Lohfink2016} except we use a more physical model \texttt{nthComp} to describe the warm corona emission. Using the same disk assumption as in \textit{Model 2}, we find two solutions to the spectra. The best-fit parameters are shown in Tab.~\ref{fits-Kerr}. The first solution (we name it \textit{Model 3.1}) gives a high spin, a steep emissivity profile, and a low inclination angle, and is broadly consistent with the best-fit model in \citet{Lohfink2016}. The second solution (\textit{Model 3.2}) gives a lower $\chi^2$ and prefers a negative spin parameter of the black hole. Fig.~\ref{models} shows the comparison of spectra components for the two solutions. The blurred reflection model in \textit{Model 3.1} is fitting some broad and featureless component. Nevertheless, in \textit{Model 3.2} it tends to fit a narrow feature near the Fe ${\rm K}\alpha$ peak. The residuals in the top panel of Fig.~\ref{ratios-Kerr} indeed shows a weak hump at 6--7 keV. Although we usually assume that the accretion disk extends to the ISCO, a truncated disk is also possible. We explore such a geometry in \textit{Model 3.3} assuming a maximally rotating black hole ($a_* = 0.998$). The other best-fit parameters shown in Tab.~\ref{fits-Kerr} are in good agreement with \textit{Model 3.2} and the inner disk radius is measured to be $13.9_{-5.2}^{+3.1}~R_{\rm ISCO}$. We show the data to model residuals of \textit{Model 3.3} with blurred reflection turned off in the bottom panel in Fig.~\ref{ratios-Kerr}. The residuals imply that the blurred reflection mainly contributes to narrow emission at 6--7 keV as well as some contribution to the flux below 2 keV.

It is worth to note that even for the best fit in Tab.~\ref{fits-Kerr}, the improvement of $\chi^2$ with respect to \textit{Model 1} is only about 40 with over 2000 degrees of freedom. The non-relativistic \textit{Model 1} already gives a good fit to the X-ray spectra of Fairall 9. In addition, we do not clearly detect the signature of a broad iron line in Fig.~\ref{ratios} and Fig.~\ref{xmme-ratio}. All these considerations lead us to conclude that the relativistic reflection component in the spectra of Fairall 9, if present, is weak. 

\begin{figure*}
    \centering
    \includegraphics[width=0.47\textwidth, trim={1.5cm 1.5cm 0.0cm 0.0cm},clip]{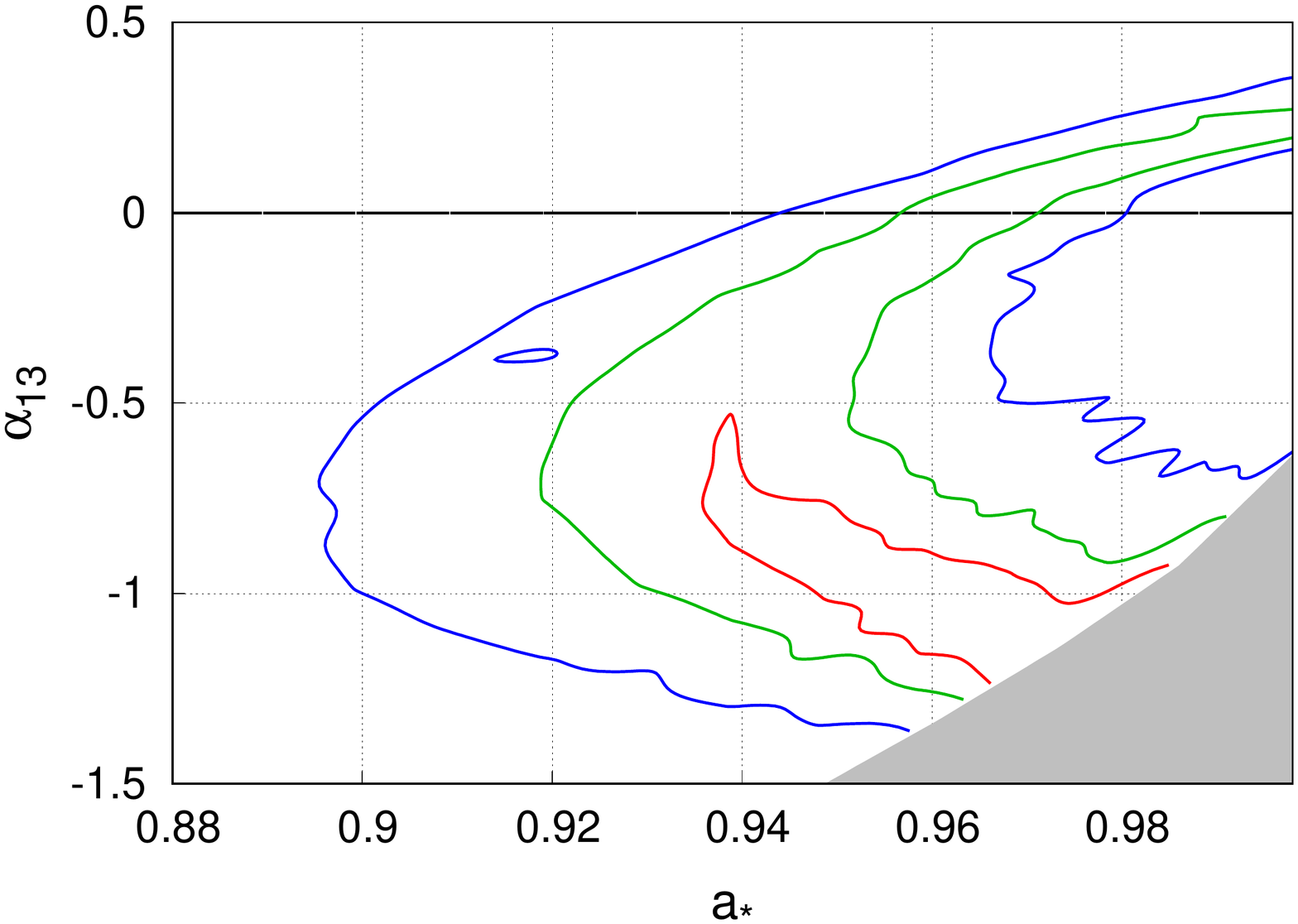}
    \includegraphics[width=0.47\textwidth, trim={1.5cm 1.5cm 0.0cm 0.0cm},clip]{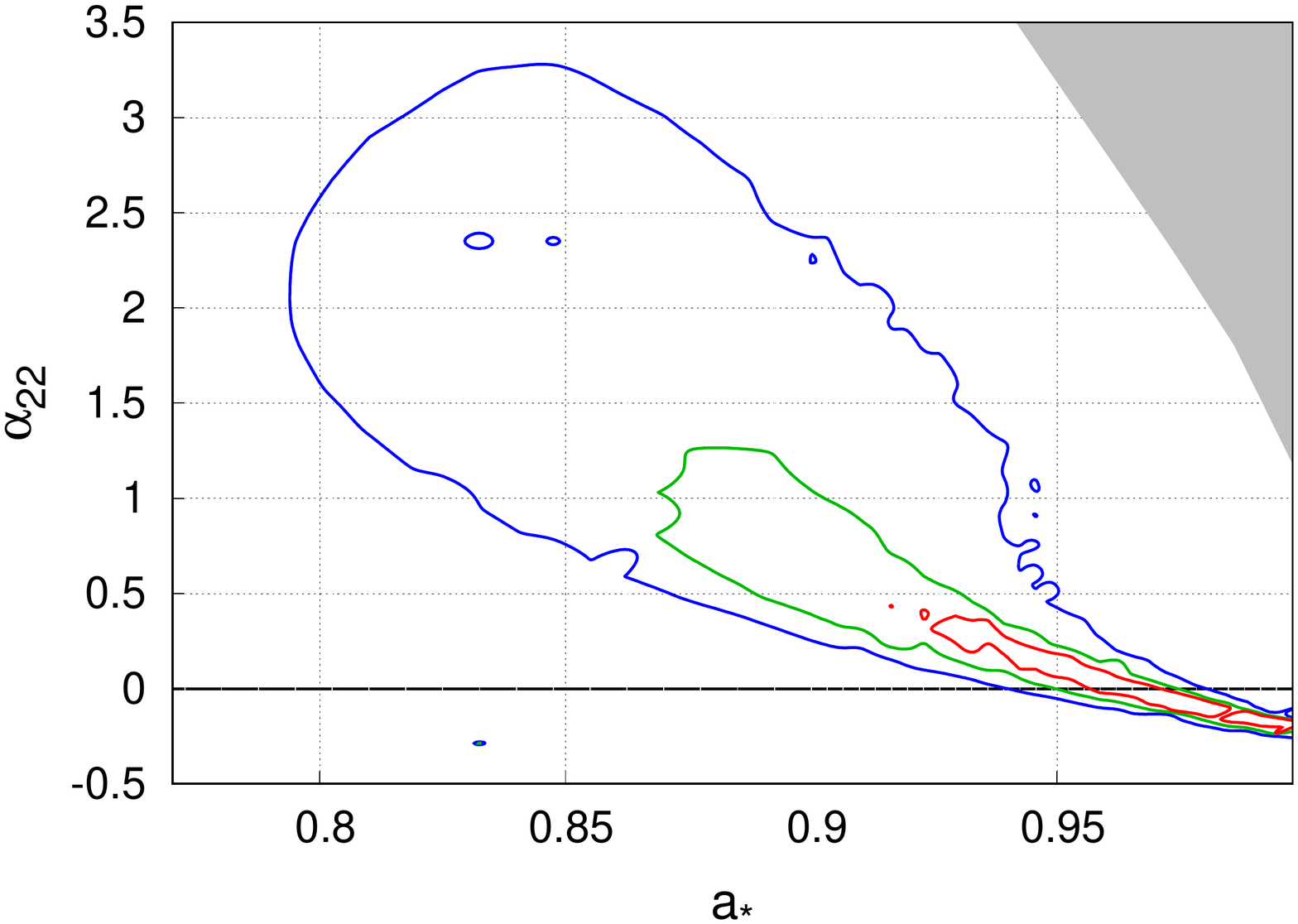}
    \caption{Confidence contour plot for spin parameter and deformation parameters. The red, green and blue lines correspond to 68\%, 90\%, and 99\% confidence levels respectively. The grey region is not included in our analysis because the spacetime there has pathological properties. \label{cont}}
\end{figure*}

To constrain possible deviation from Kerr spacetime around the black hole in Fairall 9, we replace the component \texttt{relxill} in \textit{Model 3} with its non-Kerr extension \texttt{relxill\_nk} v1.3.2~\citep{Bambi2017,2019ApJ...878...91A}. Similar to the results with \texttt{relxill}, we also find three solutions. The best-fit parameters are shown in Tab.~\ref{fits-def} and the constraints of the deformation parameters in Fig.~\ref{cont}. We only show the results corresponding to solution 1 (\textit{Model 3.1}) because the deformation parameters are totally unconstrained in the other two solutions. This is easy to understand since both negative spin and truncated disk result in a large distance between the black hole and the inner edge of the accretion disk. In this case, effect on the reflected spectrum due to deviations from Kerr background are not strong enough to be detected with current data. When assuming solution 1, we find from Fig.~\ref{cont} that the deformation parameters can recover 0, which means the Kerr solution is recovered. 

\begin{table*}
    \centering
    \caption{Best-fit values for warm corona emission and blurred reflection both included. \label{fits-Kerr}}
    \renewcommand\arraystretch{1.8}
    \begin{tabular}{lccccc}
        \hline\hline
        Description & Component & Parameter & Model 3.1 & Model 3.2 & Model 3.3 \\
        \hline
        Galactic absorption & \texttt{Tbabs} & $N_{\rm H} (10^{20} {\rm cm}^2)$ & $3.15^*$ & $3.15^*$ & $3.15^*$ \\
        \hline
        Soft excess & \texttt{nthComp} & $\Gamma$ & $2.71_{-0.08}^{+0.07}$ & $2.79_{-0.10}^{+0.05}$ & $2.77_{-0.10}^{+0.08}$ \\
        & \texttt{nthComp} & $z$ & $0.047^*$ &  $0.047^*$ &  $0.047^*$\\
        & \texttt{nthComp} & $kT_{\rm e}$ (keV) & $0.35_{-0.04}^{+0.08}$& $1.4_{-0.5}^{+0.6}$ & $1.5_{-0.3}^{+0.7}$ \\
        & \texttt{nthComp} & Norm ($10^{-3}$) & $2.0_{-0.3}^{+0.2}$ & $4.2_{-0.9}^{+1.0}$ & $4.4_{-0.9}^{+1.0}$ \\
        \hline
        Cold reflection & \texttt{xillver} & $A_{\rm Fe}$ & $2.6_{-0.4}^{+0.8}$ & $>4.4$ & $>5.0$ \\
        & \texttt{xillver} & Norm ($10^{-5}$) & $6.6_{-0.9}^{+1.0}$ & $2.1_{-0.7}^{+0.6}$ & $2.1_{-0.4}^{+0.4}$\\
        \hline
        Ionized reflection & \texttt{relxill} & $q_{\rm in} = q_{\rm out}$ & $>9.4$ & $3.1_{-0.5}^{+1.3}$ & $>3.13$ \\
        & \texttt{relxill} & $a_*$ & $0.965_{-0.014}^{+0.008}$ & $<-0.66$ & $0.998^*$ \\
        & \texttt{relxill} & $i$ (deg) & $<9.7$ & $<20.7$ & $<13.7$\\
        & \texttt{relxill} & $R_{\rm in}$ ($R_{\rm ISCO}$) & $1^*$ & $1^*$ & $13.9_{-5.2}^{+3.1}$ \\
        & \texttt{relxill} & $\log\xi$ & $3.0_{-0.7}^{+0.3}$ & $3.01_{-0.13}^{+0.15}$ & $3.02_{-0.12}^{+0.15}$ \\
        & \texttt{relxill} & Norm ($10^{-5}$) & $1.6_{-0.5}^{+0.6}$ & $0.42_{-0.10}^{+0.11}$ & $0.34_{-0.08}^{+0.12}$ \\
        \hline
        Continuum & \texttt{zCutoffpl} & $\Gamma$ & $1.93_{-0.02}^{+0.03}$ & $1.68_{-0.10}^{+0.09}$ & $1.67_{-0.16}^{+0.09}$ \\
        & \texttt{zCutoffpl} & $E_{\rm cut}$ (keV) & $1000^*$ & $120_{-34}^{+77}$ & $118_{-25}^{+54}$ \\
        & \texttt{zCutoffpl} & Norm ($10^{-3}$) & $7.6_{-0.4}^{+0.3}$ & $5.2_{-1.0}^{+0.9}$ & $5.0_{-1.0}^{+1.0}$ \\
        \hline
        &  $\chi^2/\nu$ & & 2208.63/2136 & 2187.62/2135 & 2180.82/2135 \\
        & $\chi^{2}_{\nu}$ & & 1.03 &  1.02 & 1.02 \\
        \hline
        \tablenotetext{0}{\textbf{Note.} Model 3: \texttt{Tbabs*(nthComp + relxill + xillver + zCutoffpl)}. The parameters with $^*$ are fixed at given values in the fit. Uncertainties are given at 90\% confidence level.}
    \end{tabular}
\end{table*}

\begin{table}
    \centering
    \caption{Best-fit values with model relxill\_nk. \label{fits-def}}
    \renewcommand\arraystretch{1.8}
    \begin{tabular}{cccc}
        \hline\hline
        Component & Parameter & $\alpha_{13}$ & $\alpha_{22}$ \\
        \hline
        \texttt{Tbabs} & $N_{\rm H} (10^{20} {\rm cm}^2)$ & $3.15^*$ & $3.15^*$ \\
        \hline
        \texttt{nthComp} & $\Gamma$ & $2.68_{-0.05}^{+0.07}$ & $2.69_{-0.05}^{+0.06}$ \\
        \texttt{nthComp} & $z$ & $0.047^*$ & $0.047^*$  \\
        \texttt{nthComp} & $kT_{\rm e}$ (keV) & $0.33_{-0.04}^{+0.05}$ & $0.34_{-0.04}^{+0.05}$ \\
        \texttt{nthComp} & Norm ($10^{-3}$) & $1.94_{-0.22}^{+0.04}$ & $2.0_{-0.2}^{+0.2}$ \\
        \hline
        \texttt{xillver} & $A_{\rm Fe}$ & $2.5_{-0.6}^{+0.7}$ & $2.5_{-0.7}^{+0.8}$ \\
        \texttt{xillver} & Norm ($10^{-5}$) & $6.6_{-0.7}^{+0.7}$ & $6.6_{-0.8}^{+1.0}$\\
        \hline
        \texttt{relxill\_nk} & $q_{\rm in} = q_{\rm out}$ & $>9.3$ & $>6.2$ \\
        \texttt{relxill\_nk} & $a_*$ & $0.972_{-0.05}^{+0.008}$ & $>0.96$ \\
        \texttt{relxill\_nk} & $i$ (deg) & $<6.8$ & $<9.6$ \\
        \texttt{relxill\_nk} & $\log\xi$ & $3.0_{-0.5}^{+0.8}$ & $3.0_{-0.5}^{+0.22}$ \\
        \texttt{relxill\_nk} & $\alpha_{13}$ & $<-0.3$ & - \\
        \texttt{relxill\_nk} & $\alpha_{22}$ & - & $-0.16_{-0.04}^{+0.25}$\\
        \texttt{relxill\_nk} & Norm ($10^{-5}$) & $1.52_{-0.24}^{+0.6}$ & $1.5_{-0.4}^{+0.7}$  \\
        \hline
        \texttt{zCutoffpl} & $\Gamma$ & $1.93_{-0.03}^{+0.04}$ & $1.94_{-0.02}^{+0.03}$ \\
        \texttt{zCutoffpl} & $E_{\rm cut}$ (keV) & $1000^*$ & $1000^*$ \\
        \texttt{zCutoffpl} & Norm ($10^{-3}$) & $7.6_{-1.4}^{+0.3}$ & $7.7_{-0.6}^{+0.3}$ \\
        \hline
        $\chi^2/\nu$ & & 2205.31/2135 & 2207.38/2135 \\
        $\chi^{2}_{\nu}$ & & 1.03 &  1.03\\
        \hline
        \tablenotetext{0}{\textbf{Note.} Model: \texttt{Tbabs*(nthComp + relxill\_nk + xillver + zCutoffpl)}. The parameters with $^*$ are fixed at given values in the fit. Uncertainties are given at 90\% confidence level.}
    \end{tabular}
\end{table}

\section{discussion and conclusions}\label{s-dis}

Fairall 9 has a number of interesting properties (listed at the beginning of this paper) that make it a promising source for testing the Kerr black hole hypothesis using X-ray reflection spectroscopy. Unfortunately, the relativistic reflection component, if present, is weak, which makes challenging the selection of the correct astrophysical model with the available data. However, we should stress that the ideal source with all the nice properties for testing general relativity and no bad properties probably does not exist, so we have to try to do our best with the available sources and data. Moreover, in the case of Fairall 9 we can expect (see Subsection~\ref{sssss} below) that future observations with a higher energy resolution near the iron line can solve the degeneracy and be capable of selecting the astrophysical model and the physical best-fit solution.

\subsection{General considerations}

We studied a simultaneous \textit{XMM} + \textit{NuSTAR} observation of Fairall 9. The soft excess in the spectra can be described by a Comptonization component (\textit{Model 1}) and the fit is already good. When we add a blurred reflection component and we assume that the inner edge of the disk is at the ISCO radius, we find two solutions. The first solution gives a very high spin value and a broad, featureless reflected spectrum smoothed by strong relativistic effect near the black hole. The second solution, with a slightly improved statistics, needs a negative spin and fits a narrow feature near Fe ${\rm K}\alpha$. We cannot tell which solution is preferred by current data, but there are no natural mechanisms to create retrograde disks around AGNs. Prolonged disk accretions naturally produce fast-rotating supermassive black holes, a scenario that may be supported by the observation of the average high radiative efficiency of these objects~\citep{2006ApJ...642L.111W}. The capture of small bodies and/or short episodes of accretion tend to create slow-rotating black holes~\citep{2003ApJ...585L.101H,2006MNRAS.373L..90K}. Negative spins might be created by the coalescence of two supermassive black holes, as a consequence of galaxy merger, but still we could expect that the accretion process spins the final black hole up. Both the very high spin and negative spin solutions do not improve much the statistics of the non-relativistic model (\textit{Model 1}) which implies a weak contribution from blurred reflection, if present at all. No matter in which case, we alway find a low viewing inclination angle of the disk, which is consistent with the requirement for Seyfert~1 galaxies in the unified AGN model~\citep{Urry1995}.

We also explored the scenario of a truncated disk, which is theoretically more motivated than the negative spin solution. For example, truncated disks are commonly observed in X-ray binaries in the low-hard state when the accretion luminosity is low~\citep{2018ApJ...855...61W}. Assuming a maximally rotating black hole with $a_* = 0.998$, the inner disk radius of Fairall 9 is measured to be $13.9_{-5.2}^{+3.1}~R_{\rm ISCO}$. The truncated disk scenario would be consistent with the results of \citet{2017MNRAS.466.1777P}, where the authors study the lag spectrum of Fairall 9 and find that the accretion disk may be larger than that predicted by the standard disk model. On the other hand, if the corona does not illuminate the ISCO region well, the relativistic reflection spectrum of a fast-rotating black hole may be interpreted as the spectrum of a retrograde or truncated disk~\citep{2014MNRAS.439.2307F}. In our analysis, the best-fit value of the emissivity index is stuck at the maximum value allowed in the analysis ($q = 10$), even if the uncertainty is quite large (so $q > 3.13$ in Tab.~\ref{fits-Kerr}). A high emissivity index is normally thought to be possible only in the case of a fast-rotating black hole with the inner edge of the disk close to the black hole horizon, while it is unlikely for a negative spin or a disk truncated at large radii~\citep{2012MNRAS.424..217F}, as we find in our analysis. Such a consideration would thus rule out the possibility of a truncated disk. It is thus difficult to reconcile all the results already in literature with our findings, and we can neither rule out nor confirm the truncated disk scenario. Note that a source with a truncated disk would not be a good candidate for tests of general relativity in the strong field regime, as we need a disk very close to the black hole event horizon to maximize the relativistic effects in the spectrum of the source.

A warm corona seems to be necessary to explain the soft excess in the data. \textit{Model 2} with a relativistic reflection component and without warm corona provides a definitively worse fit. \citet{2019MNRAS.489.3436J} show that the soft excess can be explained with a relativistic reflection component employing a higher disk electron density, but this can unlikely explain the soft excess in the spectrum of Fairall~9. If we fit the data with \textit{Model 2} replacing \texttt{relxill} (disk electron density fixed to $10^{15}$~cm$^{-3}$) with \texttt{relxillD} (disk electron density free to vary in the range $10^{15}$-$10^{19}$~cm$^{-3}$), we do not improve the fit. The model in \citet{2019MNRAS.489.3436J} can reach higher value of the disk electron density, but the latter is supposed to decreases as the black hole mass increases, so a very high disk electron density should not be expected in Fairall~9 because of the high black hole mass. Once we admit the presence of a warm corona, we find somewhat different warm corona temperatures, ranging from $\sim 0.3$~keV (\textit{Model 3.1}) to $\sim 1.5$~keV (\textit{Model 3.2} and \textit{Model 3.3}). While the best-fit values of different models are not completely consistent among them, all models require a reasonable warm corona temperature around 1~keV, as found in other sources~\citep{2018A&A...611A..59P,Garcia2019}.

A subtle point is the estimate of the iron abundance, $A_{\rm Fe}$. In all our fits, we get an iron abundance significantly higher than 1, and in some cases the model requires very high values of $A_{\rm Fe}$ (\textit{Model 3.2} and \textit{Model 3.3}). Physical reasonings would suggest to expect $A_{\rm Fe}$ close to 1, but it is well known that a number of sources show high, or even very high, values of the iron abundance inferred from the fit of their reflection spectra~\citep{2018ASPC..515..282G}. The actual origin of such high values of $A_{\rm Fe}$ is currently unknown. \citet{2019MNRAS.489.3436J} show that reflection spectra calculated with models with higher disk electron density can alleviate the problem, finding lower values of $A_{\rm Fe}$ when they fit the data. It has been also proposed that high iron abundances may be obtained because the direct power-law component and the power-law component illuminating the disk may be different~\citep{2015ApJ...808..122F}, because the radiative levitation of iron ions in the inner part of the accretion disk may enhance the photospheric iron abundance~\citep{2012ApJ...755...88R}, or as a result of absorption from a strong disk wind~\citep{2016MNRAS.461.3954H}.

\subsection{Testing the Kerr black hole hypothesis}

Assuming that the relativistic reflection component exists and ignoring the truncated disk scenario, we can constrain the spacetime metric around the black hole in Fairall 9 and no evidence of deviations from the Kerr solution is found. For $\alpha_{22} = 0$, our constraints on the black hole spin $a_*$ and the deformation parameter $\alpha_{13}$ are (here and in what follows we report the uncertainties at the 90\% of confidence level for two relevant parameters)
\begin{eqnarray}
a_* > 0.92 \, , \quad -1.3 < \alpha_{13} < 0.3 \, .
\end{eqnarray}
For $\alpha_{13} = 0$, we find instead
\begin{eqnarray}
a_* > 0.87 \, , \quad -0.2 < \alpha_{22} < 1.3 \, .
\end{eqnarray}

As of now, the most stringent and robust constraints on the deformation parameters $\alpha_{13}$ and $\alpha_{22}$ have been obtained from the analysis of a \textit{Suzaku} observation of the stellar-mass black hole in GRS~1915+105~\citep{2019ApJ...884..147Z,2020arXiv200309663A}
\begin{eqnarray}
&& a_* > 0.988 \, , \quad -0.25 < \alpha_{13} < 0.08 \, , \quad (\alpha_{22}=0) \, , \nonumber\\
&& a_* > 0.988 \, , \quad -0.1 < \alpha_{22} < 0.3 \, , \quad (\alpha_{13}=0) \, ,
\end{eqnarray}
and from the analysis of simultaneous \textit{XMM} + \textit{NuSTAR} observations of the supermassive black hole in MCG--6--30--15~\citep{2019ApJ...875...56T}
\begin{eqnarray}
&& 0.94 < a_* < 0.98 \, , \,\,\, -0.3 < \alpha_{13} < 0.1 \, , \,\,\, (\alpha_{22}=0) \, , \nonumber\\
&& 0.91 < a_* < 0.98 \, , \,\,\, -0.1 < \alpha_{22} < 0.8 \, , \,\,\, (\alpha_{13}=0) \, .
\end{eqnarray}
GRS~1915+105 is a black hole binary, so the temperature of the accretion disk is higher, even if the source was in the low-hard state during the \textit{Suzaku} observation, and therefore it is not clear the level of accuracy of the spectra calculated with {\sc xillver}. The analysis of MCG--6--30--15 is more challenging. The source is very variable, which suggests that the geometries of the corona and of the accretion flow near the black hole change during the observation. There are several warm absorbers along the line of sight. The accretion rate of the source seems to exceed the limit of the 30\% of the Eddington-scaled luminosity for a Novikov-Thorne disk.

Other sources/observations have provided weaker constraints and/or model dependent results. In the case of stellar-mass black holes, it is often impossible to select the correct astrophysical model and we always face the problem of the higher temperature of the accretion disk~\citep{2019PhRvD..99l3007L,2019ApJ...875...41Z}. For supermassive black holes, the main difficulties are usually related to the presence of warm absorbers, the low photon count, and the selection of sources with a thin accretion disk (even because of the large uncertainties on the black hole mass and distance); see \citet{2019arXiv190508012A} for a review on current results with supermassive black holes. In the end, it is indeed challenging to find a source and an observation suitable for testing the Kerr metric using X-ray reflection spectroscopy. The ideal source presumably does not exist, and we have thus to find a compromise among all the desirable properties that should be suitable for robust tests of Einstein's gravity.

\begin{figure}
\vspace{0.7cm}
    \centering
    \includegraphics[width=0.47\textwidth, trim={0.0cm 0.0cm 0.0cm 0.0cm},clip]{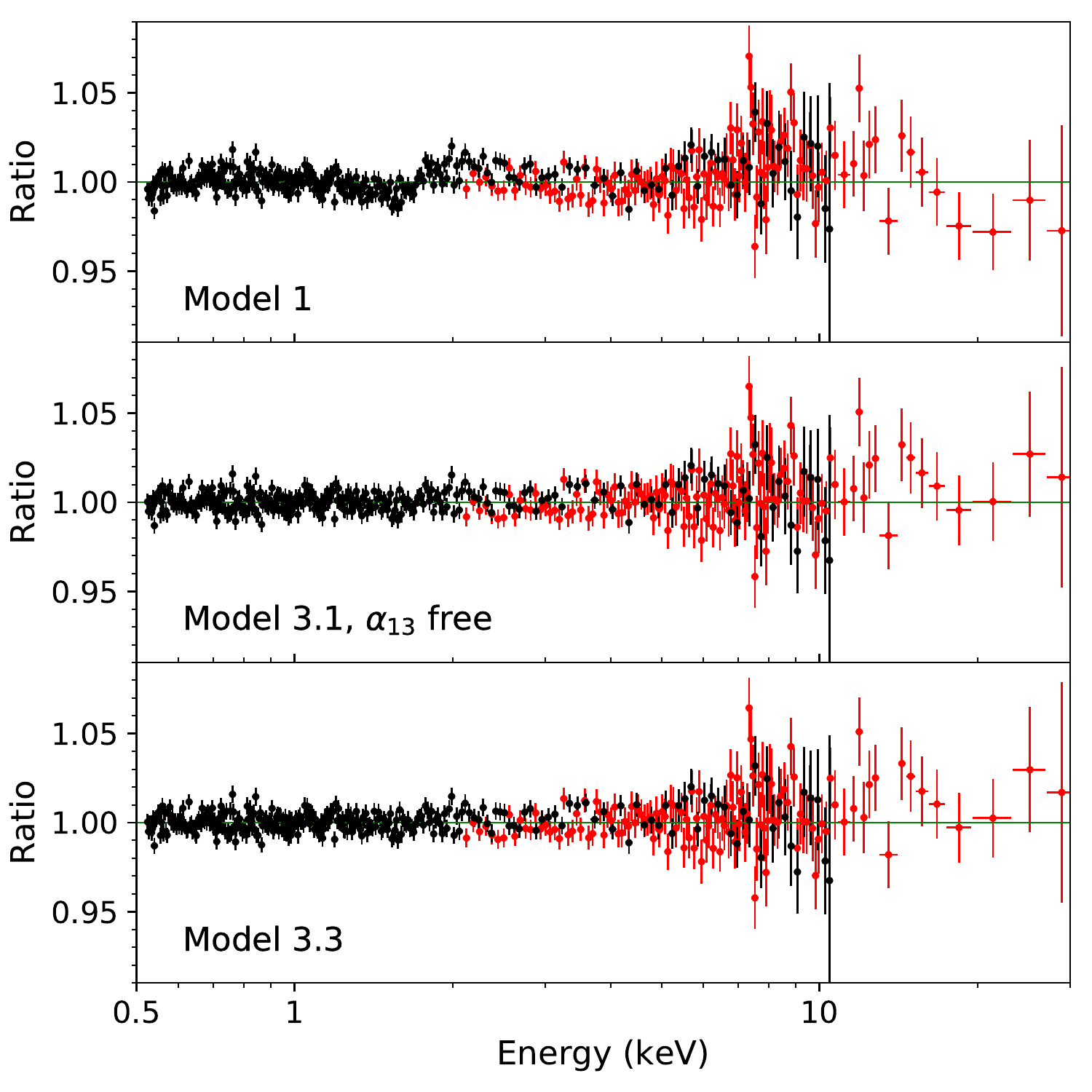}
    \caption{Data to best-fit model ratios for the simulated spectrum corresponding to Model~3.1 and fitted with Model~1, Model~3.1 with $\alpha_{13}$ free, and Model~3.3. Black data for X-IFU/\textsl{Athena} and red data for LAD/\textsl{eXTP}. \label{simu}}
\end{figure}

\begin{table*}
    \caption{Best-fit values for the simulated spectrum. \label{fits-simu}}
    \centering
    \renewcommand\arraystretch{1.8}
    \begin{tabular}{lccccc}
        \hline\hline
        Description & Component & Parameter & Model 1 & Model 3.1 & Model 3.3 \\
        \hline
        Galactic absorption & \texttt{Tbabs} & $N_{\rm H} (10^{20} {\rm cm}^2)$ & $3.15^*$ & $3.15^*$ & $3.15^*$ \\
        \hline
        Soft excess & \texttt{nthComp} & $\Gamma$ & $2.851_{-0.016}^{+0.015}$ & $2.726_{-0.03}^{+0.025}$ & $2.721_{-0.026}^{+0.020}$ \\
        & \texttt{nthComp} & $z$ & $0.047^*$ &  $0.047^*$ &  $0.047^*$\\
        & \texttt{nthComp} & $kT_{\rm e}$ (keV) & $0.447_{-0.026}^{+0.039}$& $0.355_{-0.017}^{+0.025}$ & $0.353_{-0.016}^{+0.021}$ \\
        & \texttt{nthComp} & Norm ($10^{-3}$) & $2.20_{-0.12}^{+0.19}$ & $1.93_{-0.05}^{+0.07}$ & $1.92_{-0.07}^{+0.06}$ \\
        \hline
        Cold reflection & \texttt{xillver} & $A_{\rm Fe}$ & $2.39_{-0.17}^{+0.18}$ & $2.59_{-0.16}^{+0.18}$ & $2.61_{-0.19}^{+0.17}$ \\
        & \texttt{xillver} & Norm ($10^{-5}$) & $6.6_{-0.9}^{+0.7}$ & $6.70_{-0.19}^{+0.24}$ & $6.68_{-0.28}^{+0.25}$\\
        \hline
        Ionized reflection & \texttt{relxill\_nk} & $q_{\rm in} = q_{\rm out}$ & - & $>9.6$ & $>9.0$ \\
        & \texttt{relxill\_nk} & $a_*$ & - & $>0.979$ & $0.998^*$ \\
        & \texttt{relxill\_nk} & $i$ (deg) & - & $<5.2$ & $<4.5$\\
        & \texttt{relxill\_nk} & $R_{\rm in}$ ($R_{\rm ISCO}$) & - & $1^*$ & $1.385_{-0.017}^{+0.019}$ \\
        & \texttt{relxill\_nk} & $\log\xi$ & - & $3.00_{-0.25}^{+0.08}$ & $2.99_{-0.22}^{+0.13}$ \\
        & \texttt{relxill\_nk} & $\alpha_{13}$ & - & $0.052_{-0.04}^{+0.003}$ & -\\
        & \texttt{relxill\_nk} & Norm ($10^{-5}$) & - & $1.43_{-0.11}^{+0.3}$ & $1.46_{-0.22}^{+0.20}$ \\
        \hline
        Continuum & \texttt{zCutoffpl} & $\Gamma$ & $1.909_{-0.024}^{+0.013}$ & $1.941_{-0.008}^{+0.011}$ & $1.942_{-0.009}^{+0.010}$ \\
        & \texttt{zCutoffpl} & $E_{\rm cut}$ (keV) & $>230$ & $1000^*$ & $1000^*$ \\
        & \texttt{zCutoffpl} & Norm ($10^{-3}$) & $7.72_{-0.20}^{+0.13}$ & $7.70_{-0.17}^{+0.18}$ & $7.70_{-0.14}^{+0.10}$ \\
        \hline
        &  $\chi^2/\nu$ & & 6761.1/6235 & 6633.4/6230 & 6632.6/6231 \\
        & $\chi^{2}_{\nu}$ & & 1.08 &  1.06 & 1.06 \\
        \hline
    \end{tabular}
\end{table*}

\subsection{Opportunities with future X-ray missions}\label{sssss}

Lastly, we argue that future X-ray missions have the capabilities of identifying the correct astrophysical model and thus Fairall~9 may become a valuable source for testing the Kerr black hole hypothesis. In support of this claim, we have simulated a simultaneous 100~ks observation with the X-IFU instrument (0.2-12~keV) on board of \textsl{Athena}~\citep{2013arXiv1306.2307N} and the LAD instrument (2-30~keV, achieving a total effective area $\sim 3,4$~m$^2$ in the 6-10~keV band) of \textsl{eXTP}~\citep{2016SPIE.9905E..1QZ}. A similar future observation can be seen as the counterpart of a present simultaneous observation with \textsl{XMM-Newton} and \textsl{NuSTAR}. \textsl{Athena} can provide an excellent energy resolution near the iron line ($\sim 2.5$~eV), and \textsl{eXTP} can cover a wider energy band to see the Compton hump of the reflection spectrum.

For the simulation, we choose \textit{Model~3.1}. The input values are the best-fit values in \textit{Model~3.1} and reported in Tab.~\ref{fits-Kerr}. We then fit the simulated data with \textit{Model~1}, \textit{Model~3.1} with $\alpha_{13}$ free, and \textit{Model~3.3}. The results of our fits are shown in Fig.~\ref{simu} and Tab.~\ref{fits-simu}. \textit{Model~3.3} essentially reduces to \textit{Model~3.1} because we can determine the location of the inner edge of the accretion disk with good precision. \textit{Model~1} is clearly disfavored, as it provides a higher $\chi^2$ ($\Delta\chi^2 \sim 130$) and the ratio plot shows a feature around 2~keV. The fit with \textit{Model~3.1} can recover the correct input parameters. The constraint on $\alpha_{13}$ is stronger than the best constraints obtained so far, but the key-point (irrelevant in a simulation but important for real data and a test of general relativity) would be to use a source like Fairall~9 that may limit the systematic uncertainties of the model.


\vspace{0.3cm}

{\bf Acknowledgments --}
We thank Jiachen Jiang and Riccardo Middei for useful discussions and suggestions. This work was supported by the Innovation Program of the Shanghai Municipal Education Commission, Grant No.~2019-01-07-00-07-E00035, and the National Natural Science Foundation of China (NSFC), Grant No.~11973019.

\end{document}